# Preprint Version Notice

This document is a preprint of the following publication:

**Chehade, A., Ragusa, E., Gastaldo, P., Zunino, R. (2025). Tiny Neural Networks for Session-Level Traffic Classification. In: Ruo Roch, M., Bellotti, F., Berta, R., Martina, M., Motto Ros, P. (eds) Applications in Electronics Pervading Industry, Environment and Society. ApplePies 2024. Lecture Notes in Electrical Engineering, vol 1369. Springer, Cham. https://doi.org/10.1007/978-3-031-84100-2_41**

Please cite the published version of this work as indicated above.

# Tiny Neural Networks for Session-Level Traffic Classification

Adel Chehade[1], Edoardo Ragusa[1], Paolo Gastaldo[1], and Rodolfo Zunino[1*]

DITEN, Universita` degli Studi di Genova,
adel.chehade@edu.unige.it

**Abstract.** This paper presents a system for session-level traffic classification on endpoint devices, developed using a Hardware-aware Neural Architecture Search (HW-NAS) framework. HW-NAS optimizes Convolutional Neural Network (CNN) architectures by integrating hardware constraints, ensuring efficient deployment on resource-constrained devices. Tested on the ISCX VPN-nonVPN dataset, the method achieves 97.06% accuracy while reducing parameters by over 200 times and FLOPs by nearly 4 times compared to leading models. The proposed model requires up to 15.5 times less RAM and 26.4 times fewer FLOPs than the most hardware-demanding models. This system enhances compatibility across network architectures and ensures efficient deployment on diverse hardware, making it suitable for applications like firewall policy enforcement and traffic monitoring.

## 1 Introduction

The rise in encrypted internet traffic presents significant challenges for network security, making traditional methods like Deep Packet Inspection (DPI) inadequate [1] [2]. As encryption protocols evolve, advanced traffic classification techniques are required to handle encrypted data effectively [3]. Packet classification supports various applications, including firewall enforcement, traffic monitoring, and policy-based routing [4]. Statistical and behavioral approaches using machine learning (ML) with handcrafted features have gained attention. Furthermore, Deep Neural Networks (DNNs) automate feature learning and enhance classification accuracy, but they are also resource-hungry, which is an issue when applications in the Internet of Things (IoT) are envisioned.

This paper proposes a system for session-level traffic classification using a Hardware-aware Neural Architecture Search (HW-NAS) framework. The HW-NAS framework integrates hardware constraints into the Neural Architecture

* This work was partially supported by project SERICS (PE00000014) under the MUR National Recovery and Resilience Plan funded by the European Union - NextGenerationEU

Search (NAS) process, optimizing Deep Neural Networks (DNNs) for deployment on resource-constrained devices, such as those in Internet of Things (IoT) environments. Experiments using the ISCX VPN-nonVPN dataset [5] demonstrate that the HW-NAS framework significantly reduces resource requirements while maintaining state-of-the-art classification accuracy, making it suitable for real-time applications.

The use-case scenario envisions an agent on the endpoint device, monitoring outgoing traffic and applying Artificial Intelligence (AI) models locally to detect and classify security issues. This design enables the configuration and training of ML models to balance runtime performance with resource availability [6]. Specifically, the focus is on optimizing DNNs for efficient performance on resource-constrained devices.

Traditional NAS techniques, which focus solely on accuracy, are unsuitable for resource-constrained environments. The HW-NAS framework addresses this by incorporating hardware constraints into the NAS process, facilitating the design of a model optimized for network traffic analysis on constrained devices.

This paper makes the following key contributions:

- Development of an optimized tiny DNN for deployment on constrained devices, addressing the need for efficient models in network traffic analysis.
- Comprehensive preprocessing of the ISCX VPN-nonVPN dataset, transforming raw traffic data into a format suitable for deep learning.
- Validation of the resulting model on real-world data, demonstrating its efficiency while maintaining high classification accuracy.

## 2   Related Works

Traffic classification has evolved from traditional port-based methods and DPI to ML and DNN-based approaches, driven by the increase in encrypted traffic [7]. Port-based methods struggle with port obfuscation and dynamic ports, while DPI is computationally intensive and ineffective for encrypted traffic [7].

ML approaches, such as Naïve Bayes, Support Vector Machine (SVM), and Random Forests, classify traffic based on statistical features like packet sizes and flow durations [8]. These methods require extensive feature engineering and are prone to overfitting with unbalanced data. DNN models like Convolutional Neural Networks (CNNs) and Recurrent Neural Networks (RNNs) automatically extract complex features from raw data, improving accuracy for encrypted traffic. However, they usually rely on resource-hungry architectures, which cannot be deployed on resource-constrained devices.

For session-level studies, DNN methods have been widely adopted. Wang et al. [9] used 1D-CNN and 2D-CNN on the ISCX VPN-nonVPN dataset, achieving 86.6% accuracy with 1D-CNN. He et al. [10] used gray images for CNN classification, achieving high F1 scores but missing relevant session information. Lu et al. [11] used Inception and Long Short-Term Memory (LSTM) networks, achieving over 98% accuracy. These works focus on accuracy without considering hardware constraints, making them less suitable for resource-constrained environments.

For packet-level works, Lotfollahi et al. [3] combined CNN and Stacked AutoEncoder (SAE), with the CNN model outperforming SAE. Soleymanpour et al. [12] used a cost matrix for unbalanced data, achieving high performance. However, these methods require processing large volumes of packet data, leading to long training times and high computational demands.

In mixed-level approaches, which combine session and packet levels, Cui et al. [13] used a session-packets-based model with CapsNet, showing superior performance over CNNs but with high computational costs. Seydali et al. [1] combined 1D-CNN, Bidirectional LSTM (Bi-LSTM), and Generative Adversarial Network (GAN) for data augmentation, improving performance metrics. However, their approach is computationally intensive and unsuitable for real-time scenarios, relying on handcrafted features.

Overall, existing methods are not deigned to target real-time implementation of the inference phase on resource-constrained devices. This paper proposes a HW-NAS model that tackles this issue by optimizing accuracy while minimizing hardware usage. This approach is suited for session-level classification, capturing broader traffic patterns and making it ideal for real-time applications and scalable deployment scenarios.

## 3  Methodology

Deploying DNNs on heterogeneous and constrained platforms is challenging, requiring a balance between performance and hardware requirements. HW-NAS addresses this by incorporating constraints that take into account the limitations of target devices during inference [14].

The HW-NAS framework is tailored for session-level traffic classification, focusing on systems with diverse nodes that do not rely on hardware accelerators or specialized fast memory architectures. Each node is assumed to have a processor with specific floating point operations per second (FLOPs) capability, a fixed amount of RAM, and Flash memory, which is crucial for storing model parameters, especially in devices with limited disk space.

In NAS, the search space consists of all possible candidate architectures, with the goal of finding the one that achieves the highest validation accuracy for a given dataset. Evaluating all candidate architectures is computationally infeasible, so dedicated search algorithms guide the selection process using specific evaluation criteria. HW-NAS enhances NAS by including hardware constraints, either using validation accuracy as the sole evaluation criterion or incorporating additional constraints that model the hardware requirements of the architecture.

Key constraints include the number of parameters ($|a|$), maximum tensor size ($|T|$), and (*Flops*). Parameters denote the total amount of weights in the architecture, maximum tensor size refers to the largest intermediate tensor stored during processing, and FLOPs represent the computational power needed for each inference. The optimization problem is defined as:

$$\begin{aligned} \max_{a \in A, w} \quad & \text{Accuracy}_{val}(w, a) \\ \text{s.t.} \quad & w =_w L_{train}(w, a), \\ & |a| < D_{Th}, \\ & |T| < R_{Th}, \\ & |Flops| < Flops_{Th} \end{aligned} \qquad (1)$$

Here, $w$ represents the weights trained on the training set, and $Th$ are the thresholds indicating device limits. $L_{train}$ is the training loss. Tensor values $T$ are calculated at runtime based on the input and are stored in the RAM. $R_{Th}$ represents the maximum tensor elements that can be accommodated by the available RAM. The sequential nature of DNNs allows RAM to be sized proportionally to the largest tensor. After training, network parameters $|a|$ remain constant and are stored on Flash memory, indicated by constraint $D_{Th}$. These parameters occupy the largest portion of memory needed to store a DNN. Lastly, $Flops_{Th}$ indicates the number of FLOPs per second that the target system can support. FLOPs might not be an exact measure due to possible pipelining or parallelization in multi-core processors and is considered a worst-case scenario.

The search space $A$ uses block-wise architectures incorporating 1D convolutional layers, batch normalization, activation, and optional max pooling and dropout layers. For inspecting communication data streams, 1D-CNNs are preferred as they can combine and aggregate local and global information while maintaining lower computational requirements compared to recurrent architectures or transformers [9].

Each block in the architecture can be configured with several parameters, including the number of filters, kernel size, stride value, and padding type for the convolutional layers, as well as pooling and dropout operations. Pooling operations can vary between max pooling and average pooling, with constraints to prevent errors from very small input sizes. Architectures are built by sequentially stacking these blocks.

The HW-NAS framework leverages a standard evolutionary algorithm to explore the search space [15]. This algorithm iteratively creates new candidate architectures by applying random mutations to a parent architecture. Each candidate is trained, and the one with the best evaluation result is selected as the new parent for the next iteration. This process continues until a predefined number of generations is reached.

## 4   Experimental Setup

Figure 1 provides an overview of the experimental setup, including the dataset preprocessing, HW-NAS implementation, and the optimal DNN selection.

The dataset in [5] is utilized, containing approximately 30GB of traffic data across 11 classes. This dataset includes captured traffic for various applications in pcap format, labeled according to the application and activity.

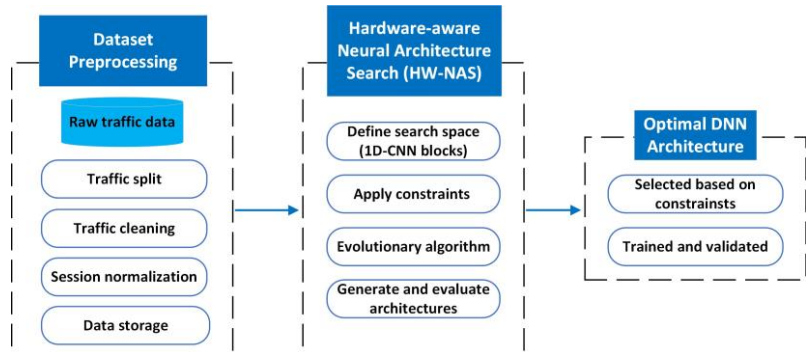

**Fig. 1.** Overview of the experimental setup

As shown in the first block of Figure 1, preprocessing involves several steps. First, raw traffic is divided into sessions using Scapy, a Python library for network packet manipulation. These sessions, which capture bidirectional traffic between the same source and destination IP addresses, ports, and protocols, are preferred over flows, which are unidirectional, for better performance in classifying encrypted traffic [9]. Next, data cleaning is performed by removing data link layer information, such as MAC addresses, and anonymizing IP addresses to prevent overfitting and ensure relevance. Packets without payloads (SYN, ACK, FIN flags) and irrelevant DNS segments are discarded. Sessions are then standardized by normalizing them to a uniform length of 784 bytes, achieved by trimming longer sessions and padding shorter ones with 0x00. Finally, the session data is scaled between 0 and 1 to ensure consistency in input values and then stored.

As depicted in the second and third blocks of Figure 1, the HW-NAS procedure is developed and executed on a workstation with a Nvidia 2080 Ti GPU. The Python code, implemented using Keras and TensorFlow libraries, automatically generates DNN architectures that meet specified constraints and trained them on the preprocessed dataset.

A validation set, comprising 20% of the training data, is extracted using a standard holdout procedure. All architectures are trained for up to 100 epochs with an initial learning rate of $10^{-3}$, a batch size of 128, learning rate reduction on the plateau, and early stopping based on validation loss. Each network is trained 5 times using a multi-start approach, and the best architecture is selected based on the validation set.

The HW-NAS executes 100 generations, each with a population of 10 candidate architectures (children), and takes approximately 4 days to complete on a Nvidia 2080 Ti GPU. The search space is defined with constraints from the literature [15]. Random mutation functions involve inserting, deleting, or modifying network blocks by changing block parameters. Hardware thresholds, including memory requirements and FLOPs, are set to the minimum values from previous studies. These measures are computed by reimplementing proposed architectures in Keras, detailed in section 5. The performance of the generated networks is evaluated using *accuracy*, *precision*, *recall* and *F1 score*.

## 5   Results

Table 1 shows a comparative analysis of session-level models from the literature. It lists for each model, respectively, the *accuracy*, *precision*, *recall*, *F1 score*, number of parameters, maximum tensor size, FLOPs, Flash memory usage, and RAM memory usage. The classification performance has been measured on a test set that has never been involved in any parameter or hyper-parameter tuning.

| Method   | Acc. (%) | Prec. (%) | Rec. (%) | F1 (%) | Params (M) | Max Tensor Size | FLOPs (M) | Flash (Mbytes) | RAM (Kbytes) |
|----------|----------|-----------|----------|--------|------------|-----------------|-----------|----------------|--------------|
| Proposal | 97.06    | 97.17     | 97.01    | 97.11  | 0.088      | 20,124          | 10.100    | 0.352          | 80.496       |
| [11]     | 98.10    | 98.00     | 98.00    | 98.10  | 19.748     | 76,248          | 41.117    | 78.992         | 304.992      |
| [16]     | 98.00    | 98.00     | 98.00    | 98.00  | 6.165      | 25,088          | 40.392    | 24.660         | 100.352      |
| [9]      | 86.60    | -         | -        | -      | 5.833      | 25,088          | 39.727    | 23.332         | 100.352      |
| [17]     | -        | -         | -        | -      | 0.223      | 313,600         | 267.217   | 0.892          | 1254.400     |
| [10]     | -        | 98.64     | 98.65    | 98.64  | 5.833      | 25,088          | 39.727    | 23.332         | 100.352      |

**Table 1.** State of the Art Model Hardware Comparison

The proposed model demonstrates competitive performance across multiple metrics: *accuracy* of 97.06%, *precision* of 97.17%, *recall* of 97.01%, and *F1 score* of 97.11%. This is achieved while significantly reducing the number of parameters and computational requirements compared to state-of-the-art models.

Table 2 further illustrates the efficiency of the proposed model by comparing both raw metrics and the ratio of each baseline model's hardware requirements to those of the proposed model. A value higher than 1 indicates that the proposed model is more efficient.

| Method   | Acc. (%) | Prec. (%) | Rec. (%) | F1 (%) | Params (Ratio) | Max Tensor (Ratio) | FLOPs (Ratio) | Flash (Ratio) | RAM (Ratio) |
|----------|----------|-----------|----------|--------|----------------|--------------------|---------------|---------------|-------------|
| Proposal | 97.06    | 97.17     | 97.01    | 97.11  | 1.00           | 1.00               | 1.00          | 1.00          | 1.00        |
| **Comparison** | | | | | | | | | |
| [11]     | 98.10    | 98.00     | 98.00    | 98.10  | 224.40         | 3.79               | 4.07          | 223.79        | 3.79        |
| [16]     | 98.00    | 98.00     | 98.00    | 98.00  | 70.05          | 1.25               | 4.00          | 69.97         | 1.25        |
| [9]      | 86.60    | -         | -        | -      | 66.28          | 1.25               | 3.93          | 66.00         | 1.25        |
| [17]     | -        | -         | -        | -      | 2.53           | 15.58              | 26.45         | 2.55          | 15.58       |
| [10]     | -        | 98.64     | 98.65    | 98.64  | 66.31          | 1.25               | 3.93          | 66.01         | 1.25        |

**Table 2.** Efficiency Comparison of the Proposed Model

The experiments confirm that the proposed method supports tight constraints while maintaining a high *accuracy*. The approach achieves an *accuracy* of 97.06%, which is higher than the results obtained by [9]. This gain is additionally supported by the fact that the number of parameters is reduced by nearly 66 times. Compared to [11] and [16], there is a slight decrease in generalization performance, but the number of parameters is reduced by over 200 times and the number of FLOPs by nearly 4 times.

A direct comparison with [17] is challenging because the original paper reports separate metrics for non-VPN and VPN data. For non-VPN data, their model achieved a *precision* of 87.6%, *recall* of 87.3%, and an *F1 score* of 87.5%; for VPN data, a *precision* of 95.2%, *recall* of 97.4%, and an *F1 score* of 96.1%. The proposed model performs better overall, offering 15.58 times less RAM usage and 26.45 times fewer FLOPs per inference. Similarly, compared to [10], the proposed model demonstrates competitive performance metrics, while being significantly more efficient in hardware usage.

In summary, the HW-NAS approach effectively generates neural network architectures for session-level traffic classification, significantly reducing hardware resource requirements while maintaining high performance.

## 6    Conclusion

This study presented a system for session-level traffic classification using the ISCX VPN-nonVPN dataset, supported by a HW-NAS approach. The resulting model demonstrated competitive performance across multiple metrics while significantly reducing hardware resource requirements compared to state-of-the-art models, making it well-suited for deployment in environments with limited computational resources.

Future work will extend this approach to packet-level traffic classification to achieve comparable or superior performance to existing state-of-the-art models while further reducing hardware requirements. This involves optimizing the HW-NAS framework to handle finer granularity of packet-level data effectively. Additionally, both session-level and packet-level models will be implemented and tested on various edge devices to evaluate their practical applicability and performance in real-world scenarios.